\documentclass[aps,prb,preprint]{revtex4}
\usepackage[T1]{fontenc}
\usepackage[latin1]{inputenc}

\makeatletter

\providecommand{\LyX}{L\kern-.1667em\lower.25em\hbox{Y}\kern-.125emX\@}
\let\SF@@footnote\footnote
\def\footnote{\ifx\protect\@typeset@protect
    \expandafter\SF@@footnote
  \else
    \expandafter\SF@gobble@opt
  \fi
}
\expandafter\def\csname SF@gobble@opt \endcsname{\@ifnextchar[
  \SF@gobble@twobracket
  \@gobble
}
\edef\SF@gobble@opt{\noexpand\protect
  \expandafter\noexpand\csname SF@gobble@opt \endcsname}
\def\SF@gobble@twobracket[#1]#2{}

\makeatother

\begin{document}

\title{Exciton states and optical properties of CdSe nanocrystals}

\author{Jes\'{u}s P\'{e}rez-Conde}

\address{Departamento de F\'{\i}sica, Universidad P\'{u}blica de Navarra, E-31006
Pamplona, Spain}

\author{A. K. Bhattacharjee}

\address{Laboratoire de Physique des Solides, UMR du CNRS, Universit\'{e} Paris-Sud,
91405 0rsay, France}

\date{\today{}}

\begin{abstract}
The optical spectra of CdSe nanocrystals up to \( 55 \) \AA{} in diameter are
analyzed in a wide range of energies from the fine structure of the low-energy
excitations to the so-called high-energy transitions. We apply a symmetry-based
method in two steps. First we take the tight-binding (TB) parameters from the
bulk \( sp^{3}s^{*} \) TB model, extended to include the spin-orbit interaction.
The full single-particle spectra are obtained from an exact diagonalization
by using a group-theoretical treatment. The electron-hole interaction is next
introduced: Both the Coulomb (direct) and exchange terms are considered. The
high-energy excitonic transitions are studied by computing the electric dipole
transition probabilities between single-particle states, while the transition
energies are obtained by taking into account the Coulomb interaction. The fine
structure of the lowest excitonic states is analyzed by including the electron-hole
exchange interaction and the wurtzite crystal-field terms in the exciton Hamiltonian.
The latter is diagonalized in the single electron-hole pair excitation subspace
of progressively increasing size until convergence. The peaks in the theoretical
transition spectra are then used to deduce the resonant and nonresonant Stokes
shifts, which are compared with their measured values in photoluminescence experiments.
We find that the final results depend on the crystal-field term, the relative
size of the surface and the degree of saturation of the dangling bonds. The
results show a satisfactory agreement with the available experimental data.
\bigskip{}
\end{abstract}
\maketitle


\section{Introduction}

The optical properties of semiconductor nanocrystals have attracted considerable
attention in recent years due to their possible applications in quantum dot
lasers and other devices (see for example Ref. \onlinecite{bg98}) and, from
a more basic point of view, because the nanocrystals are the physical realization
of small systems where the effect of the low dimensionality should be most important.
We are interested here in CdSe nanocrystals, which have been intensively studied
by several groups. \cite{mn93,er96,cg96,cd97,cd98,hc94,nb96} The high-energy
excitonic transitions have been investigated to some extent, \cite{hc94,nb96}
but most of the experimental works have been devoted to the study of the size
dependence of the photoluminescence and the fine structure of the low-energy
exciton states.\cite{er96,cg96,cd97,cd98} In particular, the observation of
the {}``dark exciton{}'' in CdSe quantum dots (QD's) is one of the more salient
features.

Several theoretical models have been proposed following two main starting points.
First, the {}``particle in the box{}'' point of view is realized in the effective
mass approximation (EMA) used by several groups \cite{er96,lx00} and the effective-bond-orbital
model by Lahel and Einevoll. \cite{le97} From the other point of view, where
the QD electron wave function is explicitly constructed from the atomic-like
orbitals we note the pseudopotential \cite{wz98,ff99} and the semiempirical
TB \cite{lp98,lw99} methods.

The TB model that we present here is particularly suitable for describing small
nanocrystals. Comparing to the EMA theories, which are known to be inadequate
for the energy gap at small sizes, some surface conditions can be varied continuously
and their influence in the final result evaluated. Also, we analyze the many-body
exciton Hamiltonian in a controlled way: the electron-hole space is expanded
until the convergence in energies is reached. The cited EMA theories, in contrast,
use a fixed number of electron-hole states. The pseudopotential calculations
published so far\cite{wz98,ff99} have been limited to smaller sizes (\( D<40 \) \AA)
than ours and some results such as the crystal-field splitting of the valence
band seem to be controversial: this splitting is expected to be smaller in QD's
than in the bulk. \cite{ef92}

Finally, the previous TB calculation of the exciton fine structure by Leung
\emph{et al.} \cite{lp98} is based on an unsatisfactory truncation procedure.
These
authors first use a Lanczos algorithm to deduce a few orbital (spin-degenerate)
single-particle states below and above the energy gap and
then introduce
the spin-orbit coupling in the restricted subspace used for diagonalizing the
exciton Hamiltonian. However, in CdSe nanocrystals,
the spin-orbit interaction
is almost an order of magnitude larger than the average spacing of the TB
single-particle levels, and drastically modifies the level scheme.
Thus their method seems inappropriate beyond the immediate
vicinity of the exciton ground state.
In this paper we propose a different approach that allows a rather
complete study of the exciton spectrum. It is based on symmetrized
single-particle states which are first obtained by an exact
diagonalization of the full zinc-blende TB Hamiltonian including
the spin-orbit interaction. The relatively small crystal-field term
 ($\sim 1/15$-th of the spin-orbit one) 
representing the wurtzite crystal structure is then diagonalized
in the subspace spanning the exciton states. Not surprisingly,
we obtain a
fine-structure spectrum of the exciton ground state quite different from
Ref. 13, especially in terms of the relative intensities of the components.
Moreover, our study is not limited to the lowest-energy
excitations: We present results on the nonresonant Stokes shifts and the high-energy
excitonic transitions.

This paper is organized as follows. In Section \ref{theory} we give a brief
description of the TB model, the single-particle states, the exciton Hamiltonian
and its diagonalization. In Section \ref{results} we compare our results with
the available experimental data and previous theoretical analyses. Finally,
in Section \ref{summary} we summarize the main results and present some concluding
remarks.

\section{Tight-binding description of the exciton states}

\label{theory}

We consider spherical zincblende crystallites of diameter ranging from \( 16.63 \) \AA{}
(\( 87 \) atoms) to \( 54.78 \) \AA{} (\( 3109 \) atoms). In Table \ref{cf}
we list some selected crystallite sizes along with their number of atoms and
number of dangling bonds. The one particle states are obtained using the \( sp^{3}s^{*} \)
nearest-neighbor TB model introduced by Vogl \emph{et al.} \cite{vh83} extended
here to include the spin-orbit interaction. We take the tight-binding parameters
for CdSe from Ref. \onlinecite{aj98}, except some minor changes: \( E_{s,a}=-11.53 \),
\( E_{p,a}=0.53 \), \( E_{s,c}=1.83 \), \( E_{p,c}=5.87 \), \( V_{s,s}=-3.07 \),
\( V_{x,x}=1.8 \), \( V_{x,y}=4.23 \), \( V_{s,p}=2.17 \), \( V_{p,s}=-5.48 \),
\( E_{s^{*},a}=7.13 \), \( E_{s^{*},c}=6.87 \), \( V_{s^{*},p}=1.99 \), \( V_{p,s^{*}}=-3.06 \)
eV. The spin-orbit couplings are \( \lambda _{Se}=0.1434 \) and \( \lambda _{Cd}=0.059 \)
eV. The dangling bonds are passivated by hydrogen atoms where the H energy level
is put to \( E_{s,H}=-3.3 \) eV following the scaling prescription given by
Kobayashi \emph{et al.} \cite{ks82} We assume that the hopping integrals between
the anion or cation and hydrogen follow the Harrison scaling rule: \( V_{b-H}=(d_{a-c}/d_{b-H})^{2}V_{ac} \)
where the \( d_{b-H} \) are the bond lengths, with \( b \) denoting cation
(\( b=c \)) or anion (\( b=a \)). The degree of saturation of the dangling
bonds is controlled by varying these bond lengths.

We follow a symmetry-based method developed previously\cite{pb99} to obtain
the one-particle states. These are computed by diagonalizing the Hamiltonian
built in a block-diagonal form, in terms of the symmetrized basis corresponding
to the \( \Gamma _{6} \), \( \Gamma _{7} \) and \( \Gamma _{8} \) double
valued representations of \( T_{d} \). This method allows us to obtain the
complete one-particle spectra. The eigenstates can be written as, \begin{equation}
\label{state}
\phi _{i}=\sum _{R,k,m}C_{R,k,m}^{i}u_{m}^{k}({\textbf {r}}-{\textbf {R}}),
\end{equation}
 where we have omitted \( b \) and the symmetry label of the eigenstate for
brevity. The \( u^{k}_{m}({\textbf {r}}) \) are the spin-orbit coupled atomic
orbitals, following the notation by Koster \emph{et al.} \cite{kd63} They are
constructed from the direct product of the standard basis \( \{s,s^{*},p_{x},p_{y},p_{z}\} \)
and the electron spin states, all referred to the <001> cubic axis. The \( s,s^{*} \)
orbitals transform as \( \Gamma _{1} \), the \( p \) orbitals as \( \Gamma _{5} \)
and the spin functions as \( \Gamma _{6} \) representations of \( T_{d} \).
The resulting spin-orbit coupled orbitals are thus given by \( k=\Gamma _{6},\Gamma _{6}^{*},\Gamma _{7}, \)
and \( \Gamma _{8} \). There are two sets corresponding to Cd and Se. The subscript
\( m \) is the row index and \( \textbf {R} \) denotes the atomic site.

The surface effects can be important in nanocrystals, as pointed out previously
\cite{le97,pb99}. So, we have analyzed here three different cases of dangling
bond saturation. First we take the bond lengths \( d_{Cd-H}=1.71 \) \AA{} and
\( d_{Se-H}=1.47 \) \AA{} which correspond to the sum of the covalent radii.
The analysis of the density of states indicates that there is a relatively important
presence of surface states near the band edges as in the case of CdTe quantum
dots.\cite{pb99} The second series is given for \( d_{Cd-H}=1.21 \) \AA{}
and \( d_{Se-H}=0.97 \) \AA{}. In this case the surface contribution has partially
disappeared from the states of physical interest. The last series of data correspond
to \( d_{Cd-H}=0.81 \) \AA{} and \( d_{Se-H}=0.57 \) \AA{}; here the contribution
of the surface states can be found several eV away from the band gap. From these
results one can assume that the shortening of the \( d_{b-H} \) can be thought
as equivalent to project out the surface states from the Hilbert space near
the band edge. The physical consequences of the dangling bond saturation are
also equivalent to the method followed in previous TB approaches where the dangling
bonds were explicitly removed.\cite{lp98,ll90} Therefore, we can consider the
atom-hydrogen bond length as an additional fitting parameter. The study of the
excitonic structure has been restricted to the last series because the final
results seem to be the closest to the measured values. Another surface effect,
the surface-to-volume ratio, is considered through the number of dangling bonds
relative to the total number of atoms (cations and anions). In Fig. \ref{fig_natny}
the distribution of the fraction \( \nu =N_{at}/N_{dangling} \) is shown, the
QD sizes analyzed in this paper are identified by closed squares. Some properties,
such as the crystal-field splitting of the valence band edge and the photoluminescence
Stokes shifts seem to be very sensitive to the relative number of dangling bonds.

In Table \ref{edgenergies} we show the size dependence of the one-particle
energies and the corresponding symmetries of the three highest valence states
for the case \( d_{Cd-H}=0.81 \), \( d_{Se-H}=0.57 \) \AA. It can be seen
that the two highest levels are close in energy to each other and well separated
from the rest of the band for any size. It is interesting to note that this
level scheme near the `band edge' fits in with the results of the multiband
EMA calculation by Richard {\emph {et al.}\cite{rl96} showing the nearly
degenerate quadruplets $1 SDD_{3/2}$ and $1 PFP_{3/2}$. Notice, however, that
neither of our TB $\Gamma_8$ quadruplets is dipole forbidden
(see Ref. \onlinecite{pb99}). This energy separation of the two
fourfold levels
from the rest of the band is important. In fact, our numerical results for the
low-energy exciton states, to be presented later, indicate that the restricted
\( e-h \) subspace spanned by these two highest valence and the lowest conduction
levels yields almost the same results as larger subspaces.

The actual crystal structure of the CdSe crystallites is wurtzite. \cite{mn93,cg96}
This is taken into account as usual by introducing an axial crystal field along
a trigonal direction \( <111> \) of the zincblende nanocrystal. This crystal-field
splits the atomic $p$ level. In the standard \(
sp^{3}s^{*} \)
basis an energy shift has been applied to the \( p_{z} \) orbitals.
\cite{lp98}
In our spin-orbit coupled atomic basis the crystal-field Hamiltonian leads to
a splitting of the atomic \( u^{\Gamma _{8}}_{m_{s}}({\textbf {r}}) \)
energy
level into two doublets of \( C_{3v} \) symmetry. The corresponding local operator
can be written as \begin{equation}
\label{jn2}
H_{cf}^{s}=\frac{D}{3}(J_{n}^{2}-\frac{5}{4})=\frac{D}{3\sqrt{3}}\left[ \begin{array}{cccc}
0 & -1-i & i & 0\\
-1+i & 0 & 0 & i\\
-i & 0 & 0 & 1+i\\
0 & -i & 1-i & 0
\end{array}\right] ,
\end{equation}
 with \( D=0.04 \) eV and \( J_{n} \) the total angular momentum in the \( <111> \)
direction, \( J_{n}=\frac{1}{\sqrt{3}}(J_{x}+J_{y}+J_{z}) \). This value
 reproduces the crystal-field splitting of the bulk valence band. The matrix elements
of \( J_{x} \), \( J_{y} \) and \( J_{z} \) are obtained in the \( u^{\Gamma _{8}}_{m} \)
basis in the standard way. \cite{ed60} We take the relation between the total
angular momentum basis \( |JM\rangle  \) and the \( u^{\Gamma _{8}}_{m} \)
from Ref. \onlinecite{nr92}.

In the basis of the zincblende one-particle states in Eq. (\ref{state}) a
matrix
element of the total crystal-field Hamiltonian is given by, \begin{equation}
\label{hcf}
\langle \phi _{i}|H_{cf}|\phi _{j}\rangle =\sum _{R,m,n
}(C_{R,\Gamma ^{8},m}^{i})^{*}\, C_{R,\Gamma ^{8},n}^{j}\langle
u_{m}^{\Gamma ^{8}}({\textbf {r}}-{\textbf {R}})|H_{cf}^{at}|u_{n}^{\Gamma
^{8}}({\textbf {r}}-{\textbf {R}}) \rangle \end{equation}
 where \( \langle u_{m}^{\Gamma ^{8}}({\textbf {r}}-{\textbf
{R}})|H^{at}_{cf}|u_{n}^{\Gamma ^{8}}({\textbf {r}}-{\textbf {R}})\rangle  \)
are the matrix elements of the local crystal field Hamiltonian in Eq. (\ref{jn2}).

To summarize, the total single-particle Hamiltonian that we use is
\begin{equation} \label{hsingle}
H_{single}=H_{0}+H_{cf},
\end{equation}
 where \( H_{cf} \) is given in Eq. (\ref{hcf}) and \( H_{0} \) is
the zincblende TB Hamiltonian including the spin-orbit interaction.
In order to calculate the crystal-field splitting,
\( \Delta _{cf} \), of the highest valence band level we diagonalize \(
H_{single} \) in the subspace of
a sufficiently large number of valence states of \(H_0\) to reach convergence.
The results for several
surface conditions are given in Table \ref{cf} where it can be seen that the
crystal-field splitting presents an irregular behavior as a function of the
QD size. We find that this behavior is related to the relative number of dangling
bonds (see Fig. \ref{fig_natny}). In Table \ref{cf} we can see that for a
given QD size the splitting increases when the hydrogen-atom bond length decreases.
This kind of behavior related to the surface is also found in the exciton fine
structure, even if the non-monotonic jumps are not so remarkable as for the
crystal-field splitting.

A complete description of the elementary excitations is made by introducing
 the Coulomb interaction between electron and hole. Let us first define the
two-particle states needed to describe the excitations above what we call the
ground state \( |g\rangle  \), describing the fully occupied valence band.
An electron-hole state can be thought as the pair obtained when one electron
from the valence band is excited above the gap. The effective hole and electron
interact through the Coulomb interaction so that we need to describe the excitation
in terms electron-hole pairs. We call these excited states excitons because
they describe the same Hilbert state as the bulk excitons when the QD size reach
the thermodynamic limit. These exciton states
can be formally written as \begin{equation}
\label{exciton}
|e\rangle =\sum _{v,c}C_{v,c}|v,c\rangle ,
\end{equation}
 where

\begin{equation}
\label{elho}
|v,c\rangle \; =a_{c}^{\dagger }a_{v}|g\rangle ,
\end{equation}
 where the \( a_{c}^{\dagger } \)(\( a_{v} \)) is the creation(annihilation)
operator for a conduction(valence) electron and \( |g\rangle  \) is the many-particle
ground state. When the Coulomb interaction is introduced the matrix elements
of the total Hamiltonian can be written in the electron-hole basis after some
algebra as, 
\begin{equation}
\label{htotal}
H_{vc,v^{\prime }c^{\prime }}=(\varepsilon _{c}-\varepsilon _{v})\delta _{vv^{\prime }}\delta _{cc^{\prime }}-J_{vc,v^{\prime }c^{\prime }}+K_{vc,v^{\prime }c^{\prime }},
\end{equation}
 with \begin{equation}
\label{coulomb}
J_{vc,v^{\prime }c^{\prime }}=\langle \phi _{c}(1)\phi _{v^{\prime }}(2)|\frac{V({\textbf {r}}-{\textbf {r}}^{\prime })}{\epsilon ({\textbf {r}}-{\textbf {r}}^{\prime })}|\phi _{v}(2)\phi _{c^{\prime }}(1)\rangle ,
\end{equation}
\begin{equation}
\label{exchange}
K_{vc,v^{\prime }c^{\prime }}=\langle \phi _{v^{\prime }}(1)\phi _{c}(2)|\frac{V({\textbf {r}}-{\textbf {r}}^{\prime })}{\epsilon ({\textbf {r}}-{\textbf {r}}^{\prime })}|\phi _{v}(2)\phi _{c^{\prime }}(1)\rangle ,
\end{equation}
 where the \( \phi _{v(c)} \) are given in Eq. (\ref{state}), \( V({\textbf {r}}-{\textbf {r}}^{\prime }) \)
is the bare Coulomb interaction and we have explicitly included the dielectric
constant \( \epsilon ({\textbf {r}}-{\textbf {r}}^{\prime }) \). In the preceding
expressions it is implicitly understood that we consider only excitations involving
a single \( e-h \) pair. The excitations containing two or more pairs and the
polarization produced by the surrounding ions are taken into account by means
of the effective screening in the Coulomb interaction, \( \epsilon ({\textbf {r}}-{\textbf {r}}^{\prime }) \).

In the Hartree-Fock formulation of the exciton problem adopted here the
dielectric constant is introduced phenomenologically without any distinction
between the Coulomb (direct) and exchange terms. On the other hand, in the
many-body
formulation in terms of the Bethe-Salpeter equation \cite{sr66},
the exchange term appears \emph {unscreened} in the GW approximation. Indeed,
Rohling and Louie\cite{rl98} have calculated
the exciton fine structure in hydrogenated Si clusters with an unscreened
exchange interaction. However, in the EMA theory of the exciton
in bulk semiconductors it was early shown\cite{rt81,kz72} that only the
short-range part
of the e-h exchange interaction remains unscreened, but the long-range part is
screened. Recently, Franceschetti \emph{et al.}\cite{fw98} have argued that
in a QD
the exchange interaction (Eq.(9)) contains an important long-range part
which needs to be screened. Indeed, they found that within their
pseudopotential theory, unscreened exchange leads to an excitonic
splitting much larger than the experimental values. As explained below,
we reach a similar conclusion in the TB model.
There
is, of course, no basic controversy over the screening of the Coulomb term.
We assume a
uniform static dielectric constant, with
a size dependence roughly following Ref. \onlinecite{wz96}.
As for the screening of the exchange term, we carried out two different
calculations: (i) completely unscreened, and (ii) unscreened up
to the nearest neighbors, but screened beyond. A comparison of our results
with the experimental Stokes shift data clearly favors the second choice.

The TB approach aims to give an appealing physical description with only a few
adjustable parameters. Following this spirit we try to reduce the number of
integrals by means of some reasonable approximations. Let us take a generic
integral in the local spin-coupled basis which appears when Eq. (\ref{coulomb})
and Eq. (\ref{exchange}) are expanded in terms of the QD states given in
Eq. (\ref{state}), 

\begin{equation}
\label{twointeg}
\langle u^{k_{1}}_{m_{1}}({\textbf {r}}-{\textbf {R}}_{1})u^{k_{2}}_{m_{2}}({\textbf {r}}^{\prime }-{\textbf {R}}_{2})|\frac{V({\textbf {r}}-{\textbf {r}}^{\prime })}{\epsilon ({\textbf {r}}-{\textbf {r}}^{\prime })}|u^{k_{3}}_{m_{3}}({\textbf {r}}^{\prime }-{\textbf {R}}_{3})u^{k_{4}}_{m_{4}}({\textbf {r}}-{\textbf {R}}_{4})\rangle
\end{equation}

 First, we retain only integrals involving up to two distinct orbitals by setting
the subscripts as either (i) 3=2 and 4=1, or (ii) 3=1 and 4=2. The two choices
correspond the Coulomb and exchange integrals, respectively. Second, if \( {\textbf {R}}_{2}\neq {\textbf {R}}_{1} \)
the Coulomb integral is approximated by the monopole-monopole term: \( V(|{\textbf {R}}_{2}-{\textbf {R}}_{1}|) \).
Finally, the on-site Coulomb integrals are simplified: \( U_{s^{*}s^{*}}=U_{s^{*}p}=U_{s^{*}s}=0 \)
and \( U_{ss}=U_{pp}=U_{sp}=U \), but \( U \) takes different values for the
cation and the anion. These Coulomb integrals are treated as phenomenological
parameters, but follow qualitatively Ref. \onlinecite{lp98} for the on-site
Coulomb and exchange integral values.

The assumed on-site Coulomb and exchange integrals are given in Table \ref{onsiteex}.
 We have checked,
however, that the final results are not strongly dependent on these integrals
when the diagonal values change by \( 1 \) eV. Finally, the Coulomb integrals
are screened in the Coulomb Hamiltonian, they are left unscreened up to the
nearest-neighbors (primitive cell of the zincblende crystal) in the exchange
term and screened otherwise. The on-site screening factor is taken to be \( 0.4 \)
and \( 0.5 \) for cations and anions respectively. The nearest-neighbor exchange
integrals, only important in the exchange Hamiltonian, are assumed to be a tenth
of the unscreened on-site integrals.

The exciton Hamiltonian in Eq. (7) is treated by a
configuration-interaction-like
method where we take as many \( e-h \) pair states as necessary to get convergence
in the final energies and energy differences. The size of the exciton space
is not a priori fixed. For the high-energy transitions, it is adequate to evaluate
the average value of the Coulomb term and we take as many as \( 48-50 \) valence
states and \( 44-46 \) conduction states, depending on the QD size. On the
other hand, the low-energy fine structure requires a diagonalization of the
full \( e-h \) Hamiltonian including the exchange term and computer time restrictions
appear; the number of valence and conduction states needed to reach convergence
is up to \( 18 \) and \( 14 \) respectively. As in previous works \cite{ff99,lp98}
the numerical convergence is faster for the energy differences than for the
energy levels. The fine structure levels are fixed when the energy difference
of the first excited level has converged to within \( 0.1 \) meV. The highest
levels analyzed in fine structure present convergence to within \( 1 \) meV.

\section{Results}

\label{results}

\subsection{Fine structure}

\label{finestructure} An early analysis of the band-edge exciton states within
the EMA was reported by Efros \textit{et al.}\cite{er96}; it has been used
as a guideline in the TB \cite{lp98} and pseudopotential \cite{ff99} calculations.
In the spherical EMA the eightfold multiplet originating from the fourfold highest
valence and the twofold lowest conduction levels is split by the electron-hole
interaction into a fivefold passive multiplet and a threefold active triplet.
Within our TB model, the \( T_{d} \) symmetry analysis leads to three exciton
levels, \[
\Gamma _{8}^{v}\times \Gamma _{6}^{c}=\Gamma ^{vc}_{3}+\Gamma ^{vc}_{4}+\Gamma ^{vc}_{5}\]
 If we call know \( H^{Z} \) the exciton Hamiltonian which presents the zincblende
symmetry (i.e. TB Hamiltonian \( H_{0} \) plus the Coulomb and exchange interaction
terms but no crystal field) then eigenenergies of these states can be written
in terms of the matrix elements of \( H^{Z} \) as,

\begin{equation}
\label{anal_ener}
\begin{array}{c}
E_{3}=\frac{1}{2}H^{Z}_{3/2,-1/2;3/2,-1/2}-H^{Z}_{3/2,-1/2;-3/2,1/2}+\frac{1}{2}H^{Z}_{-3/2,1/2;-3/2,1/2}\\
E_{4}=\frac{1}{2}H^{Z}_{-1/2,-1/2;-1/2,-1/2}+H^{Z}_{-1/2,-1/2;1/2,1/2}+\frac{1}{2}H^{Z}_{1/2,1/2;1/2,1/2}\\
E_{5}=E_{3}+2H^{Z}_{3/2,-1/2;-3/2,1/2}
\end{array},
\end{equation}
 where for the sake of simplicity \( H^{Z}_{m1,m2;m3,m4} \) stands for \( \langle v^{8}_{m1},c^{6}_{m2}|H^{Z}|v^{8}_{m3},c^{6}_{m4}\rangle  \).
We compare now the analytical result in Eq. (\ref{anal_ener}) with the numerical
diagonalization of \( H^{Z} \) which allows us to identify the symmetry of
the states. The resulting order for any size is \( E_{3}<E_{4}<E_{5} \), but
\( E_{3} \) and \( E_{4} \) are almost degenerate.

The introduction of the crystal-field term reduces the symmetry to \( C_{3v} \),
leading to a splitting of the previously threefold degenerate levels \( \Gamma ^{vc}_{4} \)
and \( \Gamma ^{vc}_{5} \). From the \( T_{d} \) to \( C_{3v} \) compatibility
table \cite{kd63} the \( \Gamma _{4}^{vc} \) level is split into \( \Gamma _{2}^{vc}+\Gamma _{3}^{vc} \)
and \( \Gamma _{5}^{vc} \) level into \( \Gamma _{1}^{vc}+\Gamma _{3}^{vc} \).
 From this result we can deduce that the \( \Gamma _{2}^{vc} \) singlet is symmetry-forbidden.
It is important to know if the rest of these first eight levels are truly active.
To elucidate this question we first project the states obtained from the diagonalization
of \( H^{Z}+H_{cf} \) onto the eight states obtained from \( H^{Z} \). It
is known from group theory that only the \( \Gamma _{5}^{vc} \) levels can
be optically active so we only need the projection of the states obtained from
the complete Hamiltonian onto these \( \Gamma _{5}^{vc} \) active states.

The resulting projections show that the lowest state, a doublet, is not active
and the two other doublets and the \( \Gamma _{1} \) singlet are symmetry-permitted.
 However, even this analysis it is not enough to conclude that the
first doublet is forbidden, it could have other contributions from active states
of higher energy. We compute then the oscillator strength (see Appendix \ref{appendix})
for the five different levels. In Table \ref{allow-forbi} the detailed results
of this analysis are shown for a QD of size \( D=40.22 \) \AA. This allows
us to confirm that the first doublet is optically forbbiden even when \( 18 \)
valence and \( 16 \) conduction states are included to build the excitonic
matrix. This result agrees with the EMA \cite{er96} and pseudopotential \cite{ff99}
calculations, but not with that of Ref. \onlinecite{lp98}. We believe this contradiction
with the previous TB calculation arises from the different procedures followed.
We include the spin-orbit interaction from the beginning and deduce the full
single-particle spectrum. Leung \textit{et al.} considered this term perturbatively
at the same level as the Coulomb interaction, applied to a small number of band-edge
states extracted by using a Lanczos algorithm.

Let us recall that an analysis of our one-particle spectrum indicates that instead
of a fourfold degenerate valence level it is more appropriate to consider two
fourfold valence levels which are very close in energy (see Table \ref{edgenergies}).
Thus a symmetry analysis of the fine structure must take into account the first
\( 16 \) exciton states instead of \( 8 \). The analysis of this very restricted
set of states gives results which are very close to the final values when the
numerical convergence is reached. The \( 16 \) exciton states are structured
in general as two repeated eightfold multiplets. It is remarkable however that
the biggest QD presents a special behavior when convergence is reached. In
this case the degeneracy pattern of the \( 16 \) states is \( 2222121121 \)
instead of the \( 2212122121 \) for the smaller sizes.

The resonant Stokes shift is normally compared with the energy difference between
the lowest optically active (bright) and the lowest forbidden (dark) exciton
states. \cite{er96,ff99,lp98} In Fig. \ref{fig_intensity} the calculated absorption
spectra for several QD's are shown. We identify the first peak with the position
of the first allowed level. The size dependence of the resulting resonant Stokes
shift is shown in Fig. \ref{fig_resshift} and compared with the available experimental
data. There is an overall agreement between the theoretical predictions and
experimental data. The data of the smallest QD's considered here present however
a peculiar behavior. The intensity in the two smallest QD's in Fig. \ref{fig_intensity}
shows that the first transition goes to zero and also that an additional splitting
of the first active state appears and eventually could produce an unclear definition
of the theoretical Stokes shift. For example, If we associate the mean value
of the two smallest peaks of the smallest QD studied to the Stokes shift we
obtain \( 28.1 \) meV. On the other hand, one could be tempted to say that
the shift measured corresponds to the energy difference between the first of
the big peaks and the dark level. This feature might explain why the experimental
values grow so much with decreasing sizes. Another interesting feature is the
distribution of oscillator strength in EMA. It was shown that in spherical and
elliptical shape QD's the oscillator strength of the first active state, \( \pm 1^{L} \),
relative to that of the \( 0^{U} \) goes to zero for decreasing QD radius and
grows with increasing QD size.\cite{er96} We find the same qualitative feature
in our model but due to the existence of more than three active states a full
comparison is not possible. In Fig. \ref{fig_resshift} there is also another
interesting aspect concerning the results for two QD's with similar sizes, \( 30.28 \)
and \( 31.6 \) \AA{} in diameter. In fact, the difference is due to one shell
of cadmium atoms. It can be seen that the shift increases for the bigger
QD, which is opposite to the overall trend of decreasing shift with increasing
size. We associate this surprising behavior to the relative number of dangling
bonds in the surface of the QD. This perhaps account for a similar behavior
of the experimental Stokes shifts. Finally, we have also carried out
calculations assuming
a completely unscreened e-h exchange interaction; the results are shown as open
triangles connected by dotted lines. The Stokes shift is
a little enhanced in small QD's, improving the agreement with
experimental data. However, in large QD's, the enhancement is too large
and does not correctly extrapolate toward the bulk exciton splitting.
We therefore conclude that, at least
from a phenomenological point of view, it is necessary to screen the exchange interaction beyond
the first neighbor, qualitatively agreement with Ref. \onlinecite{fw98}

The interpretation of the Stokes shift of the nonresonant photoluminescence
seems less unambiguous. According to Efros \textit{et al.}, it would correspond
to the energy difference between a higher active state and the lowest active
one, while Chamarro \textit{et al.} consider it to be the energy difference
between a higher active state and the dark ground state. We have adopted the
latter point of view in presenting the experimental data in Fig.~5. In the
TB model we analyze the first \( 16 \) exciton states. The resulting excited
levels, referred to the dark ground state, are shown in Figure \ref{nonres}
along with their relative intensities. The size dependence is expressed in terms
of the exciton ground state energy as abscissa. One difficulty of the comparison
between theory and experiment here is that there is no easy way to establish
an one-to-one relation between our single-QD computed values and the measured
points. First, the presence of several levels close in energy should be considered.
Also, the size distribution of the QD's must be considered as was pointed out
by Efros {\it et al}. \cite{er96} In Fig. \ref{nonres} we show again the results
for two QD's of similar size as in Fig. 4. Note the significant differences
between them, in particular, the increase in energy of transition as well as
the shifting of the absorption intensities to higher energies. Finally,
let us mention that, as expected, an unscreened e-h exchange leads to larger
theoretical
values, yielding a better agreement with experimental data. However, a fit
to the resonant Stokes shift data seems more significant. We have
therefore retained the screening of the exchange beyond the first neighbor.

\subsection{High-energy transitions}

The high-energy transitions are also analyzed in detail. First the absorption
coefficient is computed following a simplified procedure (see Appendix \ref{appendix}).
Considering the energy scale involved, we neglect the exchange term in the interaction
Hamiltonian and treat the Coulomb term as a first-order perturbation. We identify
by inspection the major peaks; in Fig. \ref{fig_intensity_h} we show the optical
absorption spectra for several QD sizes. We deduce the symmetries of the valence
and conduction states concerned as well as their relative weights when more
than one valence-to-conduction transitions is involved. Table \ref{het} shows
the results for the \( D=40.22 \) \AA{} QD. This analysis is done first for
the smaller QD, where the peaks are well separated and clearly distinguishable.
Then the same analysis is performed for the bigger QD's, where we try to follow
the size evolution of the peaks. The existence of level crossing makes it difficult
to clearly identify these peaks. For example, the second and third peaks for
the two smallest QD's are \( \Gamma _{8}^{v}\rightarrow \Gamma _{7}^{c} \)
and \( \Gamma _{8}^{v}\rightarrow \Gamma _{8}^{c} \) transitions respectively.
The two biggest QD's present a reversed order. The evolution of the higher peaks
is more complicated, due to the mixing of different transitions into each of
the peaks.

The peaks are not immediately identified with the experimental transitions but
instead in some cases we find it appropriate to group several peaks in the same
transition. The results are summarized in Fig. \ref{fig_altener_h}.

\section{Summary}

\label{summary}

We present a unified picture of the optical properties of spherical CdSe nanocrystals
based on the \( sp^{3}s^{*} \) TB model including the spin-orbit interaction
that describes the main features of the bulk semiconductor band structure. Our
group-theoretical method has allowed us to deduce the full one-particle spectra
with symmetry-classified eigenstates for crystallite sizes up to 55 \AA{} in
diameter. The dangling bonds were passivated by hydrogen atoms. The degree of
saturation of the dangling bonds strongly influences the electronic properties
of the QD's due to the presence of surface states near the band edges. The bond
lengths from the outermost cation or anion to hydrogen were used to study this
effect. The final choice of the bond lengths removes the surfaces states completely
and optimizes the Stokes shift. The wurtzite structure of CdSe is treated as
usual by introducing a crystal-field term, reducing the symmetry from \( T_{d} \)
to \( C_{3v} \).

With the single-particle states in hand, the exciton states are written in terms
 of Slater determinants, limiting us to the subspace of single electron-hole
pair excitations. The electron-hole interaction including both the direct (Coulomb)
and exchange terms is taken into account. In order to derive the fine structure
of low-energy excitations, the full exciton Hamiltonian is diagonalized in a
subspace of progressively increasing size, with as many valence and conduction
states as necessary to reach convergence. The relative electric-dipole transition
probabilities of the exciton fine-structure components were calculated numerically
and checked against a symmetry analysis using a restricted subspace. The ground
state is found to be forbidden in all cases, in agreement with the EMA and pseudopotential
calculations, but in contradiction with the previous TB calculations,\cite{lp98}
based on a Lanczos algorithm and a perturbative treatment of the spin-orbit
interaction. The energy of the first allowed peak in the theoretical absorption
spectrum yields the resonant Stokes shift of the photoluminescence. The theoretical
values agree with the measured data except for the very small QD's. The origin
of the discrepancy is probably the increasing importance of the surface effects
in the QD's when their size decreases. Indeed it can be seen that for the small
QD's (\( D<20 \) \AA{}) there are as many or more dangling bonds as atoms.
Recently Leung and Whaley \cite{lw99} studied the influence of the surface
for small QD's and found an increase of the Stokes shift after optimized surface
relaxation.

We have also investigated the nonresonant Stokes shift which is associated to
allowed states lying above the first bright level. The theoretical results are
systematically smaller than the measured values. There are probably several
reasons for the discrepancy. First we have not included the phonons in our model
which seem to be important in the analysis of the experimental data (see Ref.
\onlinecite{er96} and references therein). Also our results are given for each
QD individually, we have not taken into account the size distribution of the
sample which apparently enhances the theoretical shift. \cite{er96}Finally,
from a theoretical point of view, we find that the exciton fine structure depends
rather strongly on the surface conditions: both the surface-to-volume ratio
and the degree of saturation of the dangling bonds. Also the geometry (crystal
structure or sample shape) plays an important role, as suggested by the increase
of the Stokes shifts with increasing crystal field.

The high-energy excitonic transitions have been also studied. The comparison
of the theoretical results with the available experimental data shows a reasonable
agreement.

\acknowledgements

We wish to thank M. Chamarro for stimulating and fruitful discussions.

\appendix

\section{Absorption coefficient}

\label{appendix} The absorption coefficient \( \alpha  \) is related to \( \epsilon _{2} \),
\( \alpha \sim \omega \epsilon _{2} \),\cite{yc96} where \begin{equation}
\label{epsilon2}
\epsilon _{2}\sim \frac{{1}}{E^{2}}\sum _{f}|M_{fg}|^{2}\delta (E-E_{fg})
\end{equation}
 where \( |M_{fG}|^{2} \) is squared transition dipole matrix given by \cite{lp98}\begin{equation}
\label{transimat}
M_{fg}=\langle f|[{\textbf {r}},H]|g\rangle =-E_{f}\sum _{vc}{C}^{*}_{vc}{\langle c|{\textbf {r}}|v\rangle }
\end{equation}
 where \( E_{f} \) is the exciton energy referred to the band gap \( E_{g} \).
It is then easy to check that the absorption coefficient can be written, in
appropriate units as, 
\begin{equation}
\label{intensity}
\alpha (E)\sim \sum _{f}E^{-1}_{f}|M_{fg}|^{2}\delta (E-E_{fg})
\end{equation}
 The dipole matrix \( \langle e|{\textbf {r}}|v\rangle	\) is written in terms
of the one particles states in Eq. (\ref{state}) as, 
\begin{equation}
\label{ehdip}
\begin{array}{c}
\langle e|{\textbf {r}}|v\rangle =\sum _{R_{s},k,m,R_{s^{\prime }},k^{\prime },m^{\prime }}{C^{c}}^{*}_{R_{s^{\prime }},k^{\prime },m^{\prime }}C_{R_{s},k,m}^{v}[{\textbf {R}}_{s}\delta _{ss^{\prime }}\delta _{kk^{\prime }}\delta _{mm^{\prime }}+\\
\\
\langle u^{k^{\prime }_{s}}_{m^{\prime }_{s}}({\textbf {r}}-{\textbf {R}}_{s^{\prime }})|
\delta \textbf{r}_{s}|u^{k_{s}}_{m_{s}}({\textbf{r}}-{\textbf{R}}_{s})\rangle ],
\end{array}
\end{equation}
 where \( \delta \textbf{r}_{s}=\textbf{r}-\textbf{R}_{s} \) . We follow the
prescription given in Ref. \onlinecite{lp98} for the non-zero elements. The
spin flips are forbidden in optical transitions and we account for that by means
of the explicit expressions for the on-site and nearest-neighbor dipole matrix
elements:

\begin{equation}
\label{dipole}
\begin{array}{c}
\langle u^{6s^{*}}_{-1/2}|\delta \textbf{r}_{s}|u^{7p}_{-1/2}\rangle =i\sqrt{\frac{1}{3}}d_{1}\textbf{e}_{z}\\
\langle u^{6s^{*}}_{-1/2}|\delta \textbf{r}_{s}|u^{8p}_{-3/2}\rangle =i\sqrt{\frac{2}{3}}d_{1}\textbf{e}_{z}\\
\langle u^{6s^{*}}_{+1/2}|\delta \textbf{r}_{s}|u^{7p}_{+1/2}\rangle =-i\sqrt{\frac{1}{3}}d_{1}\textbf{e}_{z}\\
\langle u^{6s^{*}}_{-1/2}|\delta \textbf{r}_{s}|u^{8p}_{+3/2}\rangle =i\sqrt{\frac{2}{3}}d_{1}\mathbf{e}_{z}
\end{array}
\end{equation}
 and \begin{equation}
\label{dipole2}
\langle u^{k^{\prime }}_{m^{\prime }}|\delta \textbf{r}_{s}|u^{k}_{m}\rangle =4d_{2}\delta _{kk^{\prime }}\delta _{mm^{\prime }}\textbf{e}_{z}
\end{equation}
 when \( u^{k^{\prime }}_{m^{\prime }} \) and \( u^{k}_{m} \) belong to nearest-neighbors
atoms. In the last case the \( u^{k}_{m} \) are only those originated by 
 \( s,p_{x},p_{y},p_{z} \)
atomic orbitals. For  \( d_{1} \) and \( d_{2} \) we take the numerical values
given in Ref. \onlinecite{lp98} even though we are only interested in the relative
values of the intensity.

The high-energy transitions have been analyzed by means of a simplified Hamiltonian
where the exchange term has been neglected. Additionally we take only the diagonal
correction of the Coulomb term: The expression in Eq. (\ref{transimat}) is
simplified to (\( |f\rangle =|vc\rangle  \)), \begin{equation}
\label{final}
M_{fg}=\langle f|[{\textbf {r}},H]|g\rangle =-(\varepsilon _{c}-\varepsilon _{v}-J_{vc}){\langle c|{\textbf {r}}|v\rangle }.
\end{equation}

\begin{table}
\begin{tabular}{|c|c|c|c|c|c|}
 \( D \) (\AA) &
 \( N_{at} \)&
 \( N_{dangling} \)&
 \( 0.81 \) (\AA) &
 \( 1.21 \) (\AA) &
 \( 1.71 \) (\AA) \\
\hline
\( 16.63 \)&
 \( 87 \)&
 \( 76 \)&
 \( 6.7 \)&
 \( 4.3 \)&
 \( 3.1 \)\\
\hline
\( 23.29 \)&
 \( 239 \)&
 \( 196 \)&
 \( 21.0 \)&
 \( 20.7 \)&
 \( 19.9 \)\\
\hline
\( 30.28 \)&
 \( 525 \)&
 \( 276 \)&
 \( 16.3 \)&
 \( 13.7 \)&
 \( 6.1 \)\\
\hline
\( 31.60 \)&
 \( 597 \)&
 \( 324 \)&
 \( 21.0 \)&
 \( 20.6 \)&
 \( 17.6 \)\\
\hline
\( 40.22 \)&
 \( 1231 \)&
 \( 460 \)&
 \( 18.3 \)&
 \( 15.3 \)&
 \( 8.9 \)\\
\hline
\( 54.78 \)&
 \( 3109 \)&
 \( 852 \)&
 \( 13.1 \)&
 \( 11.5 \)&
 \( 7.4 \)\\
\end{tabular}
\protect

\caption{Some properties of studied QD's are shown. The first column gives the effective
diameter, the second and third show the number of atoms and dangling bonds,
respectively. In the last three columns we give the crystal-field splitting
(in meV) of the valence band for the different surface conditions.}

\label{cf}
\end{table}

\begin{table}
\begin{tabular}{|c|c|c|c|}
 Diameter &
 \( v_{1} \)&
 \( v_{2} \)&
 \( v_{3} \)\\
\hline
\( 16.63 \)&
 \( -1.034(8) \)&
 \( -1.055(8) \)&
 \( -1.163(6) \)\\
\hline
\( 23.29 \)&
 \( -0.587(8) \)&
 \( -0.672(8) \)&
 \( -0.834(6) \)\\
\hline
\( 30.28 \)&
 \( -0.436(8) \)&
 \( -0.453(8) \)&
 \( -0.601(6) \)\\
\hline
\( 31.60 \)&
 \( -0.380(8) \)&
 \( -0.413(8) \)&
 \( -0.578(6) \)\\
\hline
\( 40.22 \)&
 \( -0.264(8) \)&
 \( -0.277(8) \)&
 \( -0.408(8) \)\\
\hline
\( 54.78 \)&
 \( -0.158(8) \)&
 \( -0.165(8) \)&
 \( -0.248(8) \)\\
\end{tabular}

\caption{Values of the three highest valence levels (in eV) for different sizes. The
symmetry is also written in parenthesis. The number in parenthesis \protect\protect\( 8\protect \protect \)(\protect\protect\( 6\protect \protect \))
is short for \protect\protect\( \Gamma _{8}(\Gamma _{6})\protect \protect \).}

\label{edgenergies}
\end{table}

\begin{table}
\begin{tabular}{|c|c|c|}
&
 \( u^{\Gamma _{6}} \)&
 \( u^{\Gamma _{8}} \)\\
\hline
\( u^{\Gamma _{6}} \)&
 13(6.5) &
 1(0.5) \\
\hline
\( u^{\Gamma _{8}} \)&
 1(0.5) &
 13(6.5) \\
\end{tabular}
\protect

\caption{On-site unscreened Coulomb and exchange integrals for cation (anion) in eV.}

\label{onsiteex}
\end{table}

\begin{table}
\begin{tabular}{|c|c|c|}
 Energy (eV) &
 \( \Gamma _{5} \) fraction &
 Oscillator Strength \\
\hline
\( 2.3076(3) \)&
 \( 0.0 \)&
 \( 0.0 \)\\
\hline
\( 2.3129(3) \)&
 \( 52.47 \)&
 \( 1.58 \)\\
\hline
\( 2.3283(2) \)&
 \( 0.0 \)&
 \( 0.0 \)\\
\hline
\( 2.3307(3) \)&
 \( 43.52 \)&
 \( 1.02 \)\\
\hline
\( 2.3351(3) \)&
 \( 97.29 \)&
 \( 1.04 \) \\
\hline
\end{tabular}
\protect

\caption{Fine structure analysis of the first eight states of the \protect\protect\( D=40.22\protect \protect \) \AA{}
quantum QD. The first column shows the energies and the corresponding symmetry
(in parenthesis). The second column shows the relative fraction of \protect\protect\( \Gamma _{5}^{vc}\protect \protect \)
state (see text) and in the third column we put the numerically calculated oscillator strength
(in arbitrary units).}

\label{allow-forbi}
\end{table}

\begin{table}
\begin{tabular}{|c|c|c|}
 Peak number &
 valence-conduction &
 Fraction \\
\hline
\( 1 \)&
 \( 8-6 \)&
 \( 1 \)\\
\hline
\( 2 \)&
 \( 8-6 \)&
 \( 1 \)\\
\hline
\( 3 \)&
 \( 7-6 \)&
 \( 1 \)\\
\hline
\( 4 \)&
 \( 8-8 \)(\( 8-7 \)) &
 \( 0.65(0.34) \)\\
\hline
\( 5 \)&
 \( 7-6 \)(\( 8-6 \)) &
 \( 0.99(0.01) \)\\
\hline
\( 6 \)&
 \( 8-8 \)(\( 6-7 \)) &
 \( 0.77(0.17) \)\\
\hline
\( 7 \)&
 \( 6-7 \)(\( 6-8 \)) &
 \( 0.75(0.25) \) \\
\hline
\end{tabular}
\protect

\caption{High energy transitions for the \protect\protect\( D=40.22\protect \protect \) \AA{}
QD. The symmetry structure of the seven first peaks are shown. If more than
a transition is concerned the two most important fractions are written.}

\label{het}
\end{table}

\protect

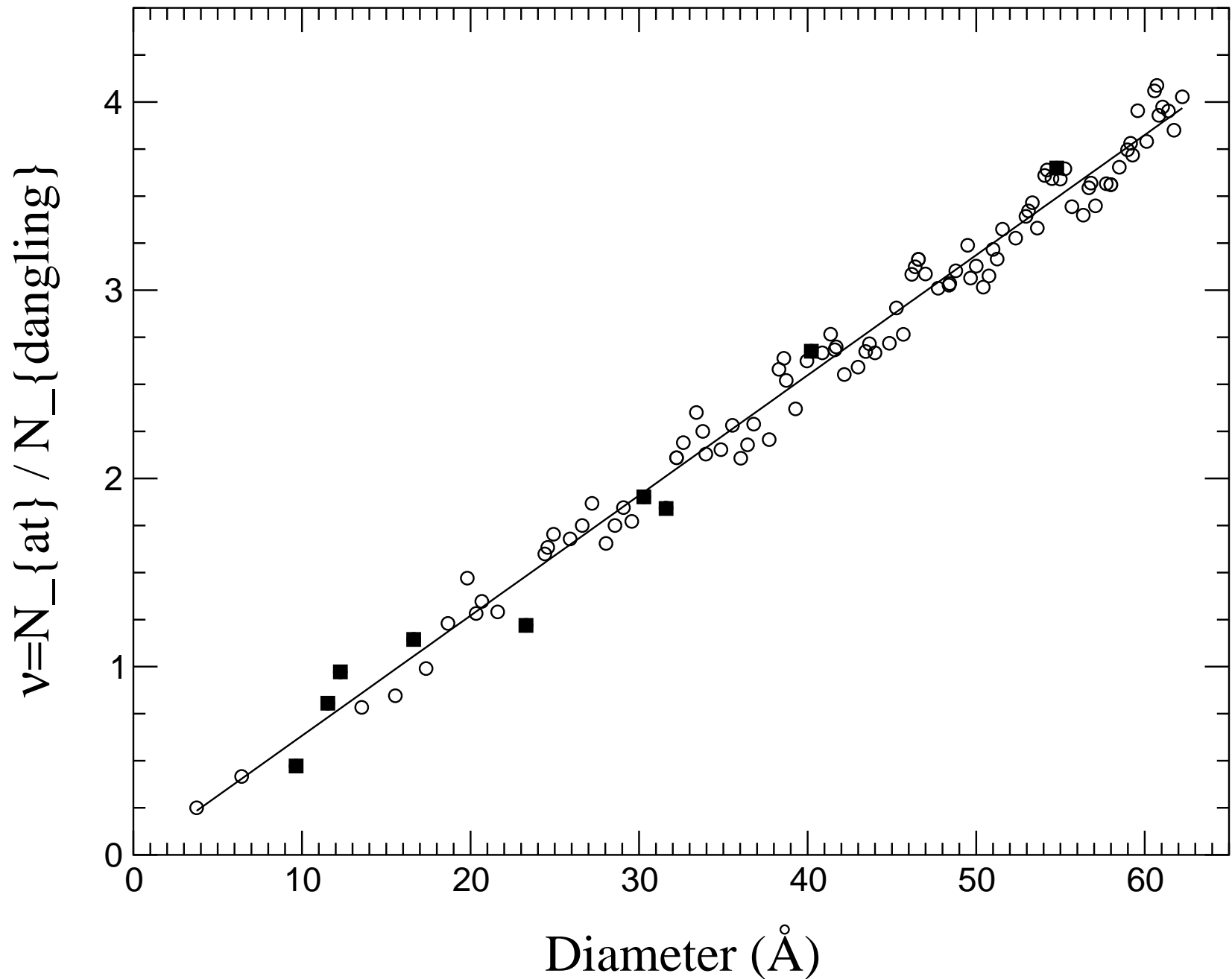
\begin{figure}

\caption{Ratio of the number of semiconductor atoms and the number of dangling bonds
(open circles). The line is the least square linear approximation. The QD's
studied in this article are shown by closed squares.}

\label{fig_natny}
\end{figure}

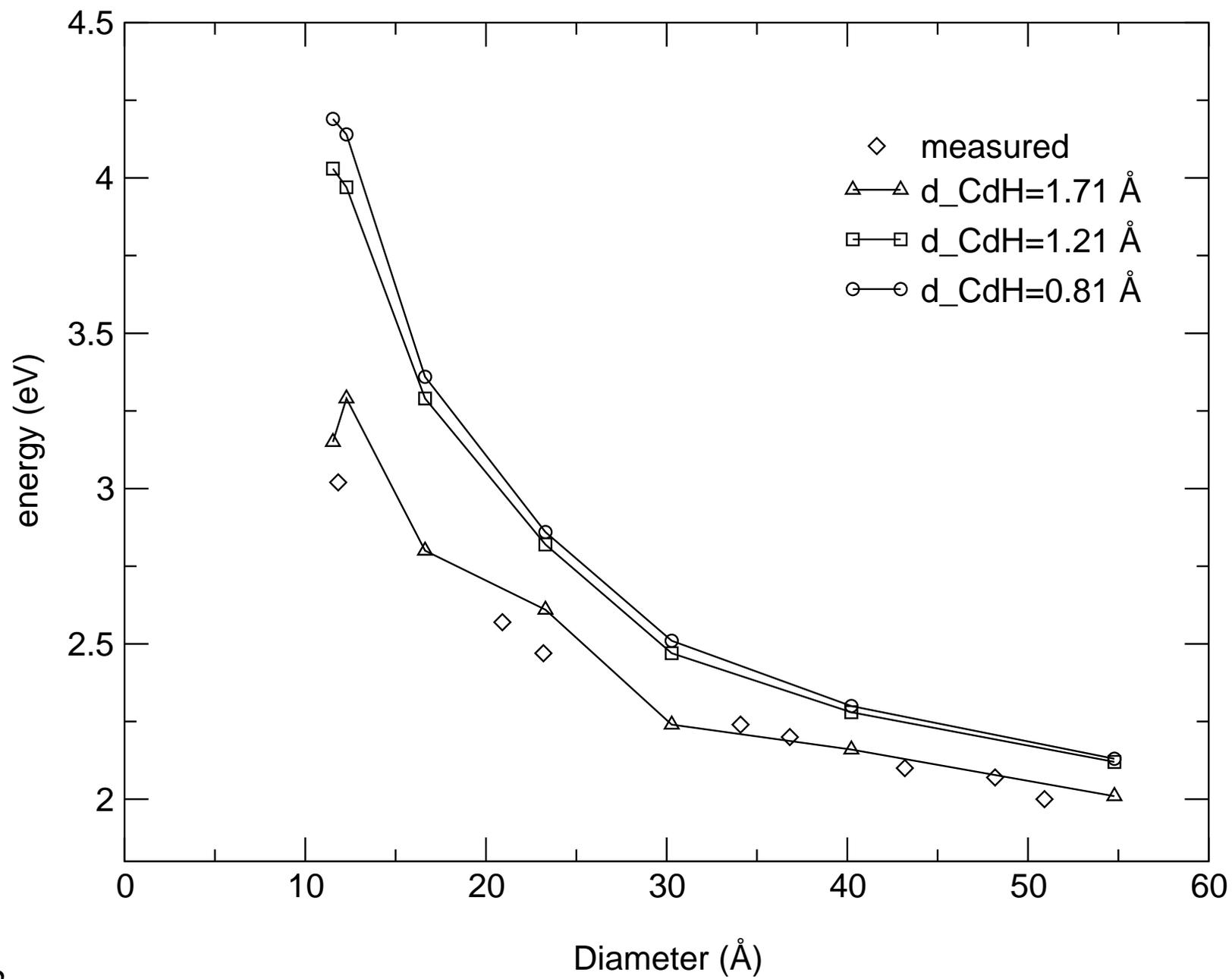
\begin{figure}

\caption{The calculated excitonic gap versus QD diameter is compared with the experimental
data from Ref. \protect\onlinecite{mn93} (diamonds). The three theoretical curves
correspond to different degrees of saturation of the dangling bonds, represented
by three different sets of cation- and anion-to-hydrogen bond lengths (see text).}

\label{gapex}
\end{figure}

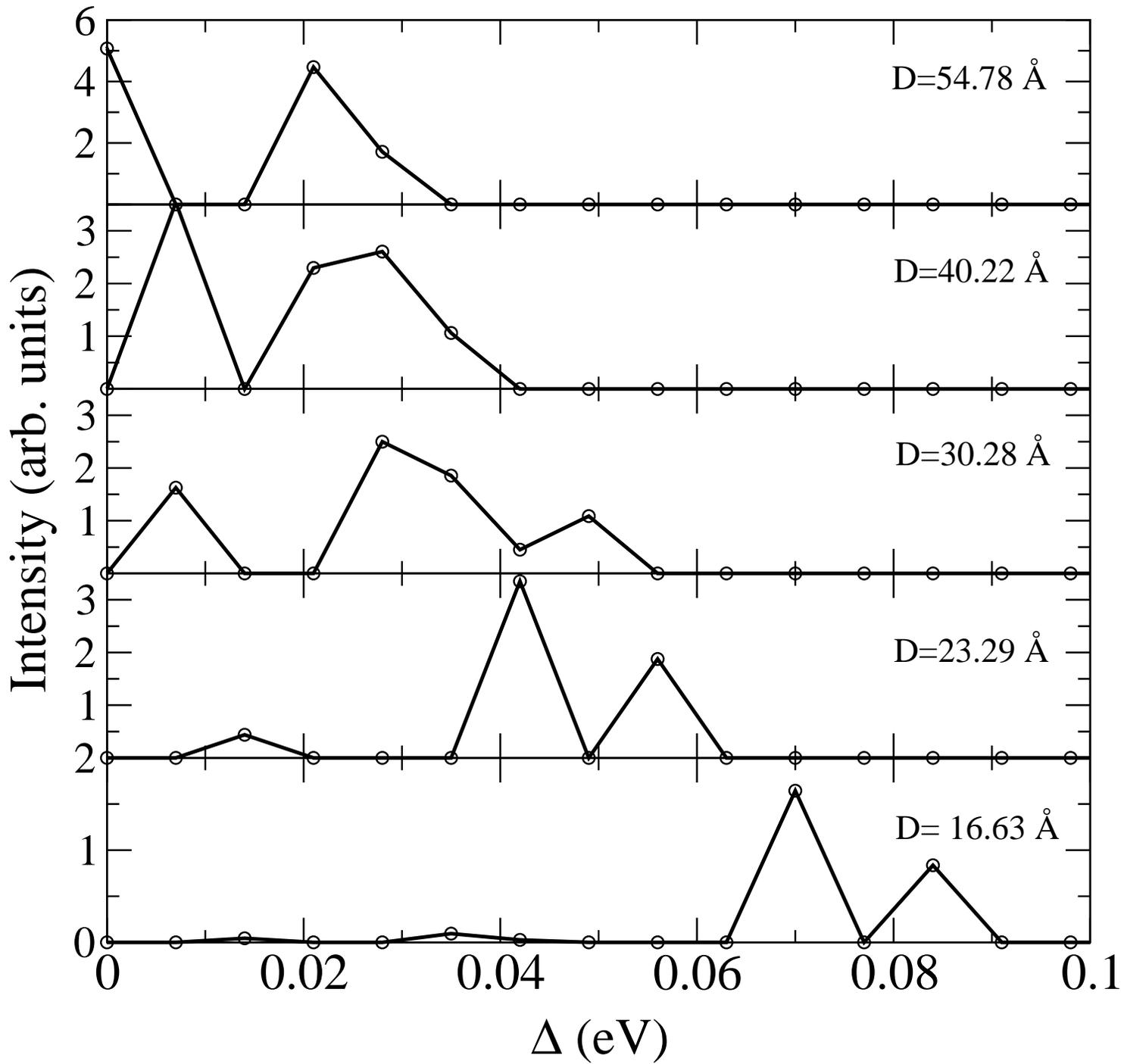
\begin{figure}

\caption{Fine structure optical absorption spectra for several sizes of CdSe nanocrystals.
The evolution of the first peak, associated to the Stokes shift shows a singular
behavior for the smallest of the QD's studied here. The size of the two lowest
peaks is negligible compared with the next peak. This behavior may be related
to the sudden increase of the measured Stokes shift in small-size QD's (see
text).}

\label{fig_intensity}
\end{figure}

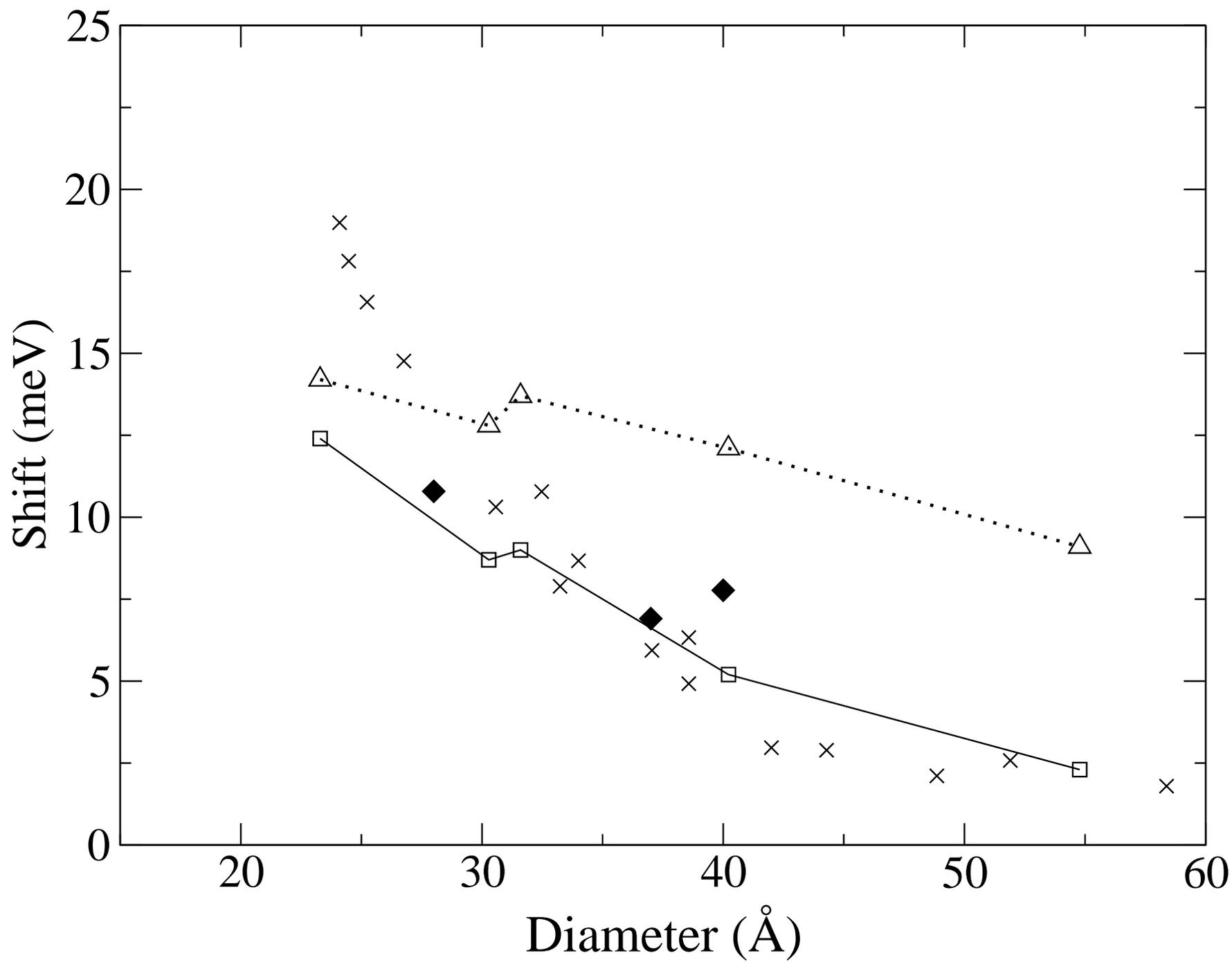
\begin{figure}

\caption{Resonant Stokes shift obtained from the energy position of the lowest energy
peak in Fig. \ref{fig_intensity} (squares), compared with the experimental
measurements from Efros \emph{et al.} \protect\cite{er96}(crosses) and Chamarro
\emph{et al.}\protect\cite{cd97}(closed diamonds). The triangles
connected by dotted lines correspond to a completely unscreened e-h exchange
interaction, while the squares connected by solid lines correspond to a finite
screening beyond the first neighbors.}

\label{fig_resshift}
\end{figure}

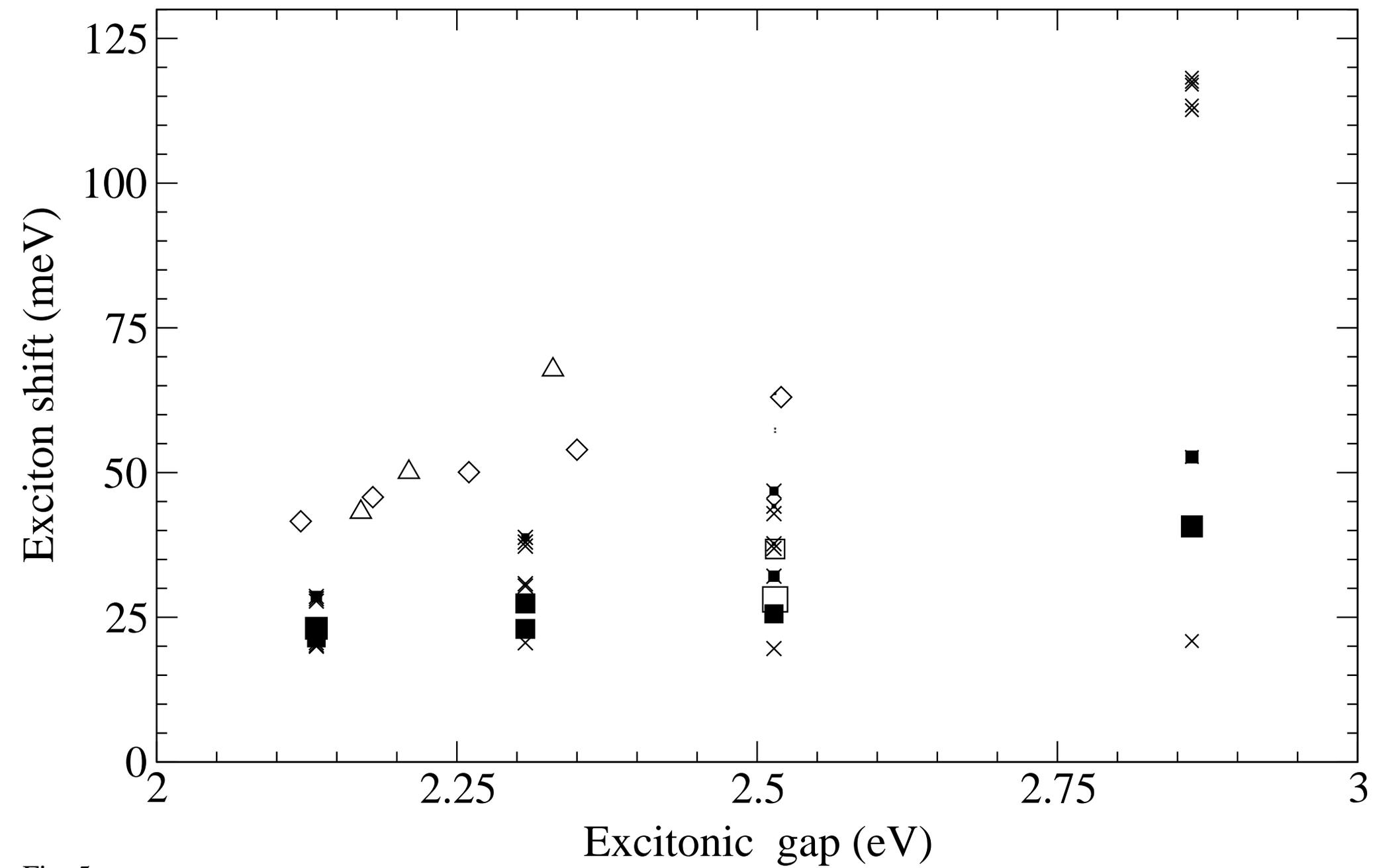
\begin{figure}

\caption{
 The size dependence of the first excitonic levels (we have omitted the first
 active level, considered before). Note that the abscissa is the exciton ground-state
 energy, instead of the QD size. All the levels are indicated by a cross. The
 allowed transitions present additionally a closed square whose size (area) is
proportional to the absorption intensity. The open squares stand for the 
\protect\protect\( D=31.60\protect \protect \) \AA{}
case.\label{nonres} We also show the measured values of the nonresonant Stokes
shifts from Norris and Bawendi \protect\cite{nb96} (open diamonds) and Chamarro
\emph{et al.} \protect\cite{cd97} (open triangles).}
\end{figure}

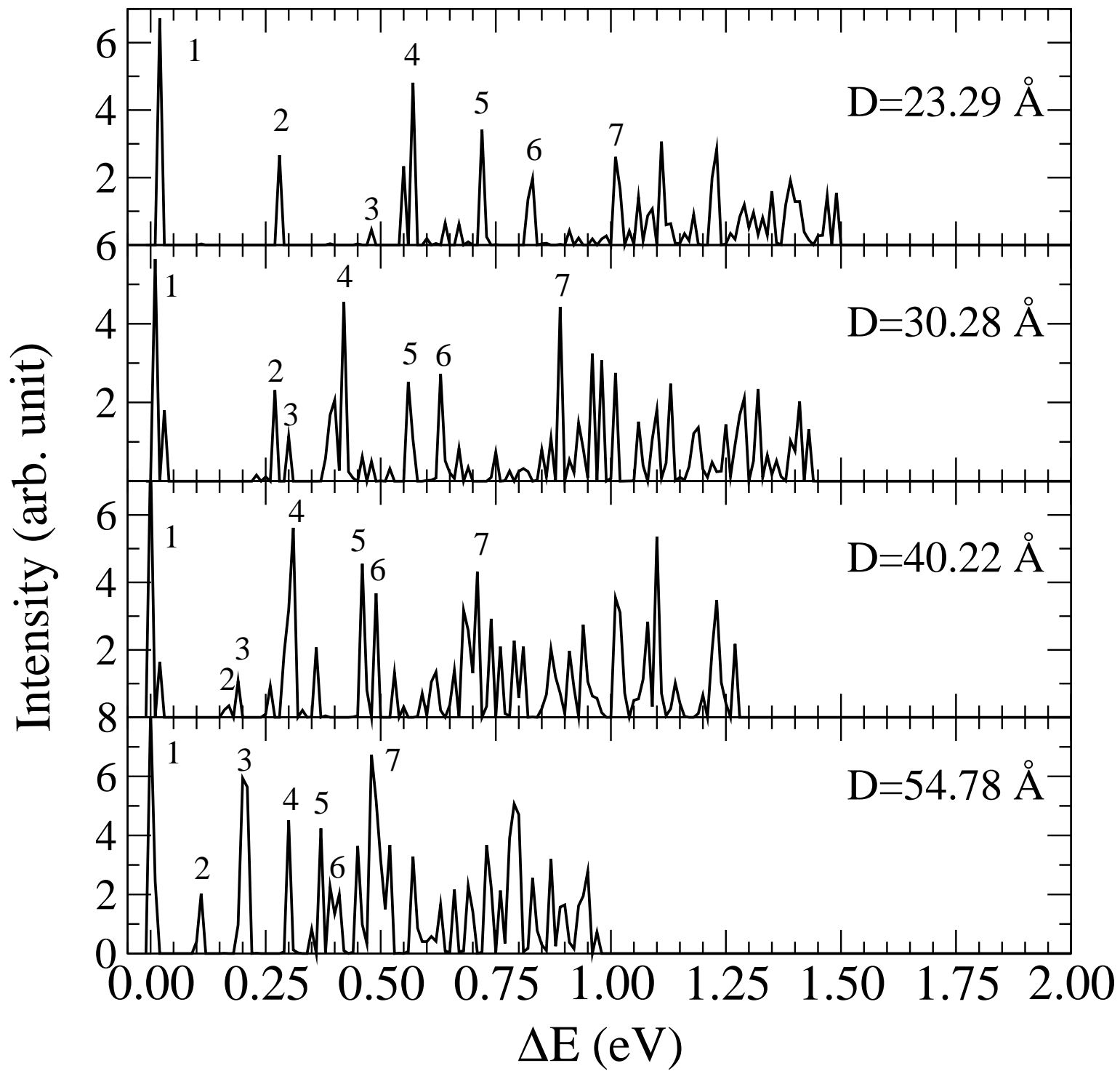
\begin{figure}

\caption{The size dependence of the absorption coefficient \protect\protect\( \alpha (E)\protect \protect \)
at high energies. The energy is measured from the `band gap' in each case.}

\label{fig_intensity_h}
\end{figure}

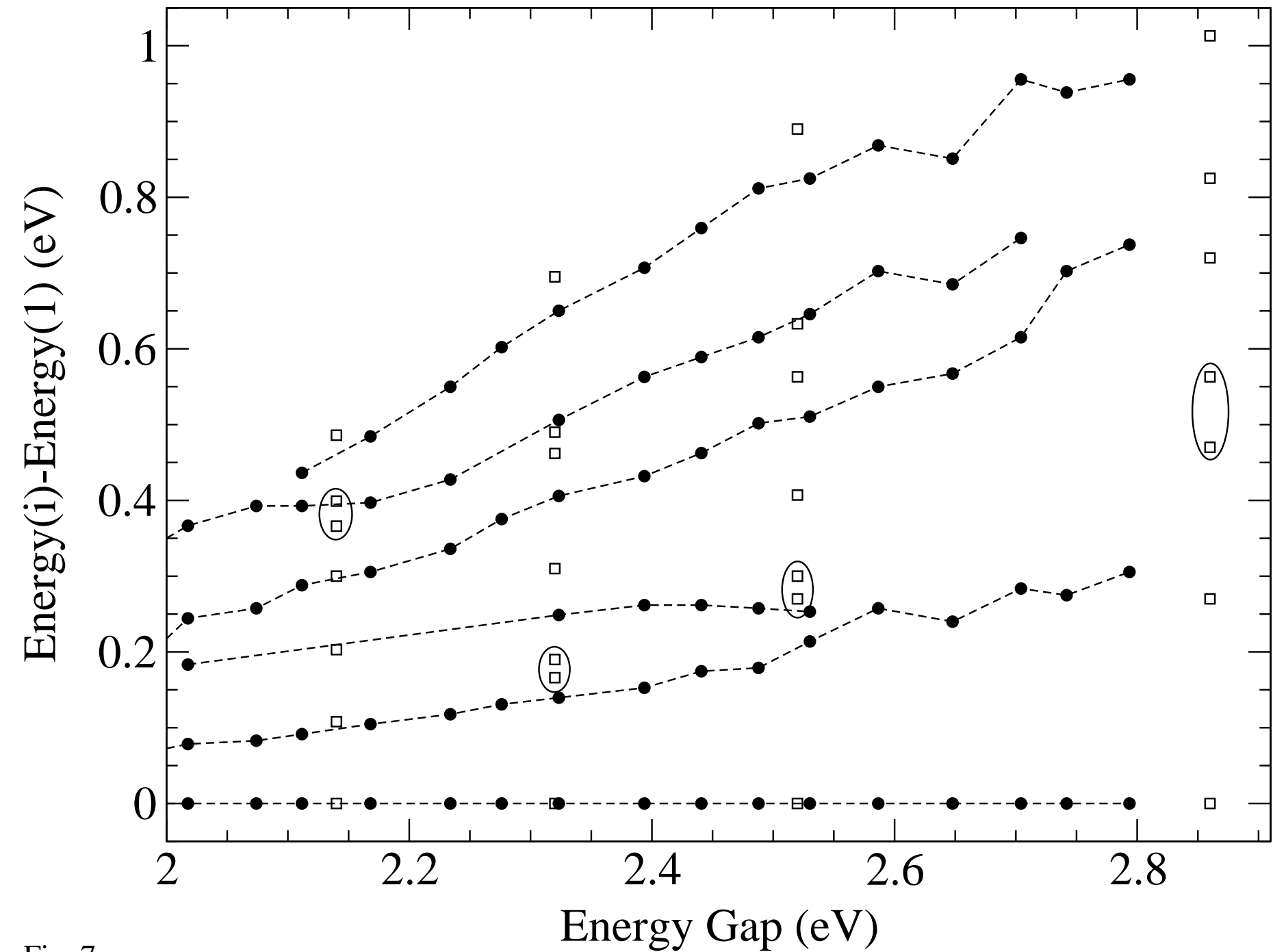
\begin{figure}

\caption{High-energy excitonic transition energies are plotted against the the excitonic
gap (open squares) and compared with the experimental data from Ref. \protect\onlinecite{nb96}
(closed circles). The encircled points suggest a probable merger.}

\label{fig_altener_h}
\end{figure}

\end{document}